\documentclass[conference, 10pt]{IEEEtran}
\IEEEoverridecommandlockouts
\usepackage{cite}
\usepackage{amsmath,amssymb,amsfonts}
\usepackage{algorithmic}
\usepackage{graphicx}
\usepackage{textcomp}
\usepackage{xcolor}
\usepackage[a4paper, total={184mm,239mm}]{geometry}

\usepackage{algorithm}
\usepackage{algorithmic}

\def\BibTeX{{\rm B\kern-.05em{\sc i\kern-.025em b}\kern-.08em
    T\kern-.1667em\lower.7ex\hbox{E}\kern-.125emX}}
\begin{document}
\bstctlcite{IEEEexample:BSTcontrol}

\title{CAIMS: Scalable Cross-Array In-Memory Search}


\author{
\IEEEauthorblockN{Yi-Chun Liao}
\IEEEauthorblockA{
\textit{Duke University}\\
Durham, NC, USA \\
yi-chun.liao@duke.edu}
\and
\IEEEauthorblockN{Liu Liu}
\IEEEauthorblockA{
\textit{University of Notre Dame} \\
Notre Dame, IN, USA \\
lliu24@nd.edu}
\and
\IEEEauthorblockN{Ruiyang Qin}
\IEEEauthorblockA{
\textit{Villanova University} \\
Villanova, PA, USA \\
qin@villanova.edu}
\and
\IEEEauthorblockN{Yiran Chen}
\IEEEauthorblockA{
\textit{Duke University}\\
Durham, NC, USA \\
yiran.chen@duke.edu}
\and
\IEEEauthorblockN{X. Sharon Hu}
\IEEEauthorblockA{
\textit{University of Notre Dame} \\
Notre Dame, IN, USA \\
shu@nd.edu}
}

\maketitle

\begin{abstract}
Search, especially top-\textit{k} similarity search, plays a vital role in many applications, but its performance can be severely limited by memory access overhead. In-memory search (IMS), an alternative to CPU/GPU-based approaches, exploits massive parallelism within a content-addressable memory (CAM) array to perform search in situ and provides both low latency and high energy efficiency. However, the capacity of a single CAM array is often insufficient for modern workloads, requiring data to be distributed across multiple arrays. Deploying scalable top-\textit{k} similarity search across multiple CAM arrays poses a fundamental problem that existing designs fail to address: how to accurately \textit{and} efficiently reduce local binary match results to global top-\textit{k} results. This work introduces CAIMS, a cross-array IMS framework that is agnostic to memory technologies and applications, enabling scalable and efficient top-\textit{k} similarity search while preserving accuracy. To enable accurate  top-\textit{k} result reduction, CAIMS leverages the existing \textit{threshold match} functionality with a combined offline and online parameter selection approach to transform local binary match results that lack globally comparable similarity information into global coarse-grained similarity scores. To further improve efficiency, CAIMS conducts hierarchical reduction and transfers only the selection results and necessary control signals with lightweight local peripherals. Implemented on a 3D NAND Flash model and demonstrated using the retrieval phase of retrieval-augmented generation, CAIMS improves throughput by up to $22.3\times/10.1\times$ and energy efficiency by $426.6\times/23.7\times$ compared to the server-class CPU baseline and prior 3D NAND Flash accelerators, respectively, while maintaining comparable accuracy.
\end{abstract}

\section{introduction}
Search, especially top-\textit{k} similarity search, is a fundamental operation for many modern applications, such as recommender systems \cite{RecommenderSystem}, memory-augmented neural networks (MANNs) \cite{MANN}, retrieval-augmented generation (RAG) \cite{RAG}, and memory-based deep reinforcement learning (DRL) \cite{DRL}. While search involves only simple computation, the overhead of massive data movements severely limits overall system performance. Previous work reveals that for data-intensive workloads, off-chip memory access can account for up to 90\% of overall latency and energy consumption \cite{offchip}, indicating that data movement is the bottleneck for search-based data-intensive applications.

In-memory search (IMS) offers a promising solution to this bottleneck. Unlike conventional CPU/GPU implementations, IMS exploits massive parallelism within a specialized memory architecture, i.e., content-addressable memory (CAM) \cite{CAMSurvey}, to perform in-situ similarity search against all candidate vectors stored within a CAM array, and produce match results based on the predefined matching scheme. By directly processing data where it resides, IMS avoids transferring the entire dataset across the memory hierarchy, producing match results in a few cycles with minimal energy consumption \cite{48Layers}.

However, although CAM efficiently identifies matches within a single array, modern workloads often exceed both CAM array depth (i.e., number of entries), which limits the number of candidate vectors, and array width (i.e., number of cells per entry), which limits the dimensionality of each candidate vector. Vector datasets must therefore be partitioned and distributed across multiple arrays along the entry and/or dimension axes, with each array producing local binary match results (match or not match). While previous work has identified this challenge \cite{CAMASim}, existing efforts support scaling along only one of the two axes \cite{HDC, ACRONYM}. Although digitizing similarity scores in each array then combining the results globally is possible, the required high-resolution ADCs can consume excessive chip area and more than 75\% of the overall energy in analog in-memory computing architectures \cite{RAELLA}. Therefore, an approach that supports scalable cross-array IMS by directly leveraging local binary match results is highly desirable.

The fundamental challenge of scalable cross-array IMS is accurate and efficient result reduction, comprising aggregation, which counts local binary match results as per-candidate scores, and selection, which identifies the global top-\textit{k} results. Regarding accuracy, each local match result lacks globally comparable similarity information. Consequently, simple aggregation does not yield faithful scores for final selection. Regarding efficiency, aggregation incurs substantial data movement across the memory hierarchy, eroding the efficiency benefit of IMS.


\begin{figure*}[t]
    \centering
    \includegraphics[width=\linewidth]{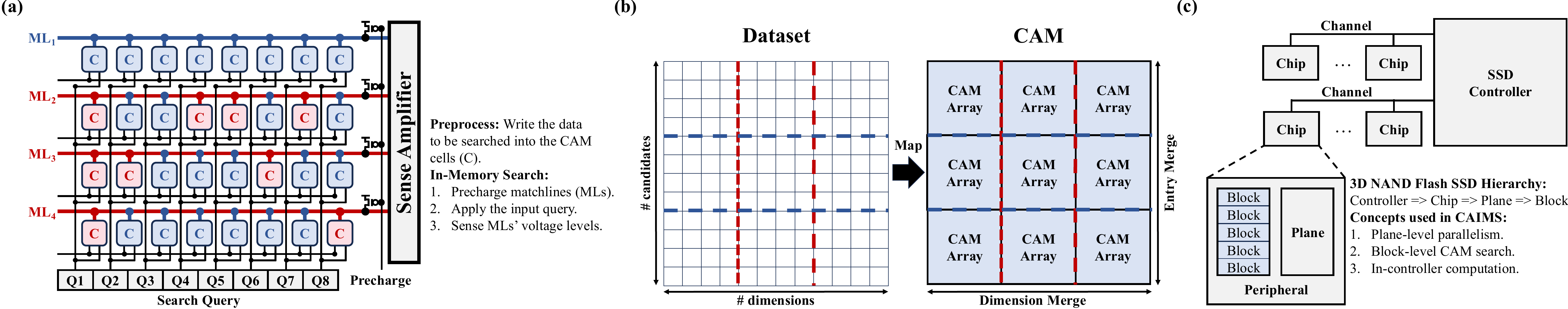}
    \caption{(a) Generic CAM array structure. (b) Entry/Dimension merge. (c) Typical architecture of a 3D NAND Flash SSD.}
    \label{fig:Background}
\end{figure*}

To address these challenges, we introduce CAIMS, a cross-array IMS framework that is agnostic to memory technologies and applications, enabling scalable and efficient top-\textit{k} similarity search while preserving accuracy (i.e., HitRate@\textit{k}). CAIMS encompasses the following novel schemes:
\begin{itemize}
    \item \textit{\textbf{Dataset-Assisted Scalable Search (DASS)}}, which combines \textit{threshold match} with offline parameter selection and aggregation to reduce local binary match results to global coarse-grained similarity scores.
    \item \textit{\textbf{Online Result Count Adaptation (ORCA)}}, which tunes the search parameter of \textit{DASS} online, adapts to query distribution shifts, and selects on average \textit{k} candidates for each query using the coarse-grained similarity scores.
    \item \textit{\textbf{Local Aggregation and Selective Transfer (LAST)}}, which performs local reduction, transferring only the selection results and necessary control signals with lightweight peripherals, thereby minimizing data movement.
\end{itemize}
To the best of our knowledge, \textbf{CAIMS} is the first fully scalable cross-array IMS design, enabling accurate and efficient top-\textit{k} similarity search. We validate CAIMS by implementing it with a 3D NAND Flash model and evaluating it on RAG retrieval. CAIMS improves throughput by up to $22.3\times/10.1\times$ and energy efficiency by $426.6\times/23.7\times$ compared to the server-class CPU baseline and prior 3D NAND Flash accelerators, respectively, while maintaining comparable accuracy.

\section{background and motivation}

\subsection{Content-Addressable Memory (CAM)}
Unlike conventional memory that retrieves data based on a given address, CAM \cite{CAMSurvey} is a memory architecture capable of comparing a query vector with all candidate vectors stored in the same array in parallel, returning the addresses of the matched results. Fig.~\ref{fig:Background}(a) shows the generic architecture of a CAM array. Data in the array are organized row-wise, with each memory cell storing the value for one dimension, and each entry corresponds to a multi-dimensional candidate vector. The query vector is applied along the dimensions and compared with candidate vectors in all entries simultaneously. A candidate is considered a match if the matchline voltage level, as detected by the associated sense amplifier, satisfies the matching criterion. 

CAM supports diverse data representations (e.g., binary \cite{CAMformer}, ternary \cite{TAPCAM}, multi-bit \cite{HDC}, and analog \cite{Analog}), distance functions (e.g., Hamming distance \cite{TAPCAM}, squared Euclidean distance \cite{SquaredEuclideanDistance}, and cosine similarity \cite{CosineSimilarity}), and matching schemes (e.g., \textit{exact match} \cite{TAPCAM}, \textit{best match} \cite{NNSinMCAM}, and \textit{threshold match} \cite{TAPCAM}). Its parallel search capability accelerates workloads such as tree-based models \cite{RETENTION}, HDC \cite{HDC}, MANN \cite{CAMMANN}, and DRL \cite{CAMDRL}.

However, device and circuit constraints inherently limit CAM array size. Prior work reports that for 4-bit per cell, extending the array width beyond 64 cells is challenging \cite{Analog}. As the number and/or dimensionality of candidate vectors grows, data must be partitioned along the entry and/or dimension axes, requiring entry and dimension merges, as shown in Fig.~\ref{fig:Background}(b).

To realize scalable cross-array IMS while preserving accuracy, entry and dimension merges should be jointly handled. Nevertheless, prior designs support merging along only one of the two axes, limiting their ability to simultaneously scale dataset size and vector dimensionality. For example, previous work \cite{HDC} applies per-array \textit{best match} and \textit{majority voting} to implement cross-array \textit{best match}. However, this approach addresses only dimension merge and assumes that every array contains the complete candidate set, limiting scaling along the entry axis. Moreover, this approach cannot be directly extended to top-\textit{k} similarity search. On the other hand, ACRONYM \cite{ACRONYM} partitions each candidate vector into two subvectors corresponding to two disjoint dimension partitions, applying \textit{threshold match} on both partitions sequentially to reduce the candidate count from $\#candidates$ to $k_{coarse}$ and finally $k_{final}$. Consequently, ACRONYM extends the vector dimension from $array\ width$ to $2\times array\ width$ with degraded recall. However, the threshold selection algorithm for \textit{threshold match}, which controls the number of matches, is not discussed. While ACRONYM mentions potential configurations to further extend the vector dimension, the practical realization is not discussed. A few other works implement cross-array IMS \cite{Macronix1, Macronix2, 48Layers}. Nevertheless, different dimension partitions are mapped to the same array as different entries, leading to matches of irrelevant query and candidate subvectors and may degrading accuracy. Hence, none of them achieves scalable cross-array IMS.

\subsection{Retrieval-Augmented Generation (RAG)}
RAG is an approach designed to extend the knowledge of large language models (LLMs) beyond training data while enhancing reliability. Typical RAG systems consist of a retriever, which compares a query vector against candidate vectors to identify the top-\textit{k} most relevant documents, and a generator, which produces the final response based on both the query and the retrieved documents. Because the retriever must search a potentially massive vector dataset for every query, retrieval can become a major performance bottleneck. Previous work \cite{RAGasService} reports that retrieval accounts for, on average, 61\% of the total runtime, while IKS \cite{IKS} reports that this proportion can reach 97\% in the worst case. Since retrieval is fundamentally a top-\textit{k} similarity search operation, RAG provides a representative workload for evaluating scalable IMS frameworks.

\subsection{3D NAND Flash SSD}
Fig.~\ref{fig:Background}(c) presents the typical architecture of a 3D NAND Flash SSD, comprising an SSD controller and hierarchically organized channels, chips, planes, and blocks. Channels, chips, and planes provide parallelism, whereas blocks within a plane share peripheral circuitry and cannot operate independently. Recent studies show that blocks can serve as CAM arrays \cite{Macronix1, Macronix2} and that the SSD controller can support general-purpose computation \cite{REIS, TVLSI26}. Combined with the high storage density of 3D NAND Flash, SSD is an ideal platform for scalable IMS.

\section{caims framework}
This section presents CAIMS, a cross-array IMS framework that is agnostic to memory technologies and applications. We first identify obstacles to accurately and efficiently reducing local binary results to global top-\textit{k} results when the dataset must be split into multiple CAM arrays. We then present an overview of the CAIMS framework and introduce the detailed algorithm-hardware co-design approach.
We finally discuss RAG retrieval acceleration using 3D NAND Flash as a case study.

\begin{figure}[t]
    \centering
    \includegraphics[width=1.0\linewidth]{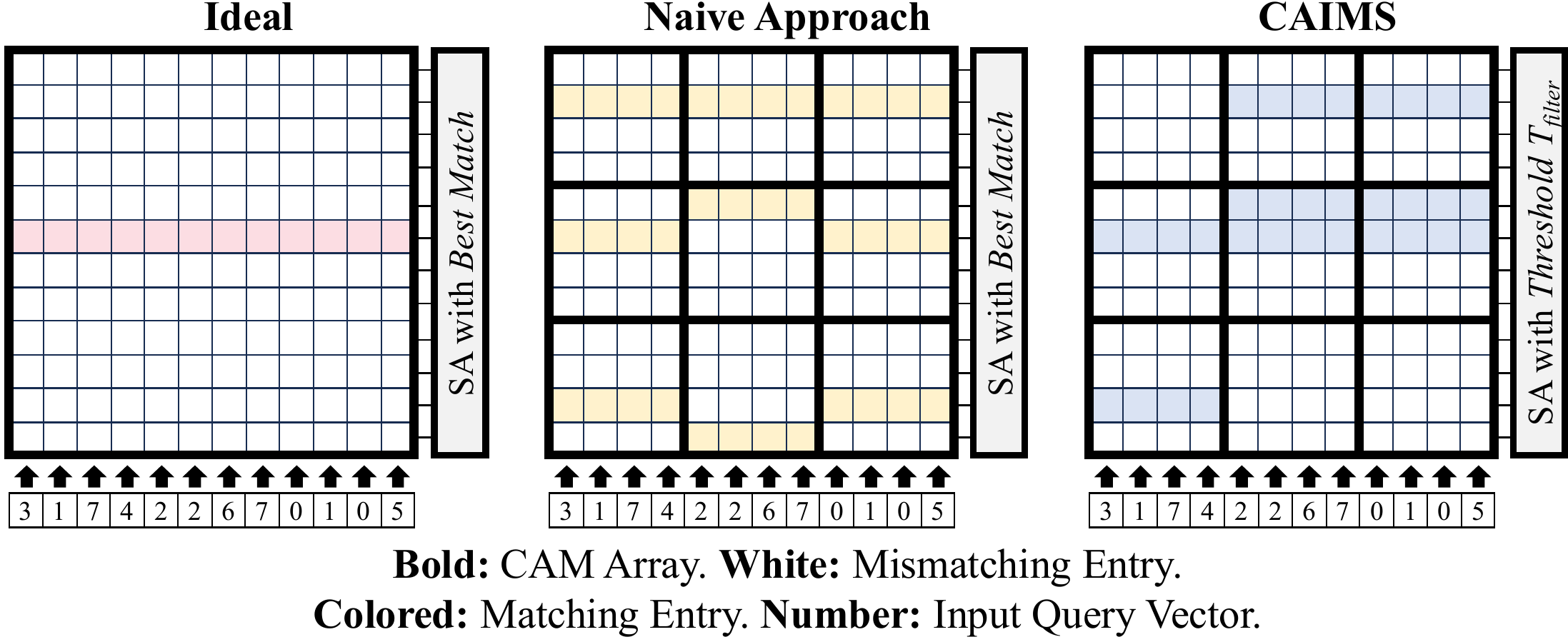}
    \caption{Comparing the ideal top-1 search result with the result obtained by a naive approach (per-array \textit{best match}) and CAIMS approach.}
    \label{fig:LocalInformation}
\end{figure}

\subsection{Overview}
The fundamental problem of scalable top-\textit{k} similarity search with IMS is: how to reduce local binary match results to global top-\textit{k} results accurately (i.e., select the most similar candidates) \textit{and} efficiently (i.e., with lower latency and energy). 

\textbf{Accuracy:} This challenge arises because local binary match results may not have the globally comparable information required for top-\textit{k} selection. Fig.~\ref{fig:LocalInformation} illustrates this issue using a top-1 example. Ideally, a sufficiently large CAM array accommodates the entire dataset and directly identifies the best match for an input query. 
However, when the dataset is partitioned across multiple arrays, each array identifies its local winner based only on the subvectors it contains. Local winners therefore indicate relative rank within an array rather than a shared level of similarity across arrays. Consequently, selecting the candidate with the most local wins can yield an incorrect global result, as illustrated by the Naive Approach in Fig.~\ref{fig:LocalInformation}.

\textbf{Efficiency:} This challenge stems from the data movement of result reduction. While IMS avoids transferring the entire $O(\#candidates\times\#dimensions)$ dataset to the CPU/GPU, centralized aggregation and selection would still require transferring $O(\#candidates\times\frac{\#dimensions}{array\ width})$ local match results. Although local aggregation and selection can reduce this traffic to $O(k\times\frac{\#candidates}{array\ depth})$, the required local selectors may consume considerable area, and many locally selected candidates may ultimately be discarded when reducing local selection results to global selection results. Efficient reduction therefore requires coordinating local and global selection to minimize unnecessary candidate transfers while keeping hardware overhead low.

\begin{figure}[t]
    \centering
    \includegraphics[width=1.0\linewidth]{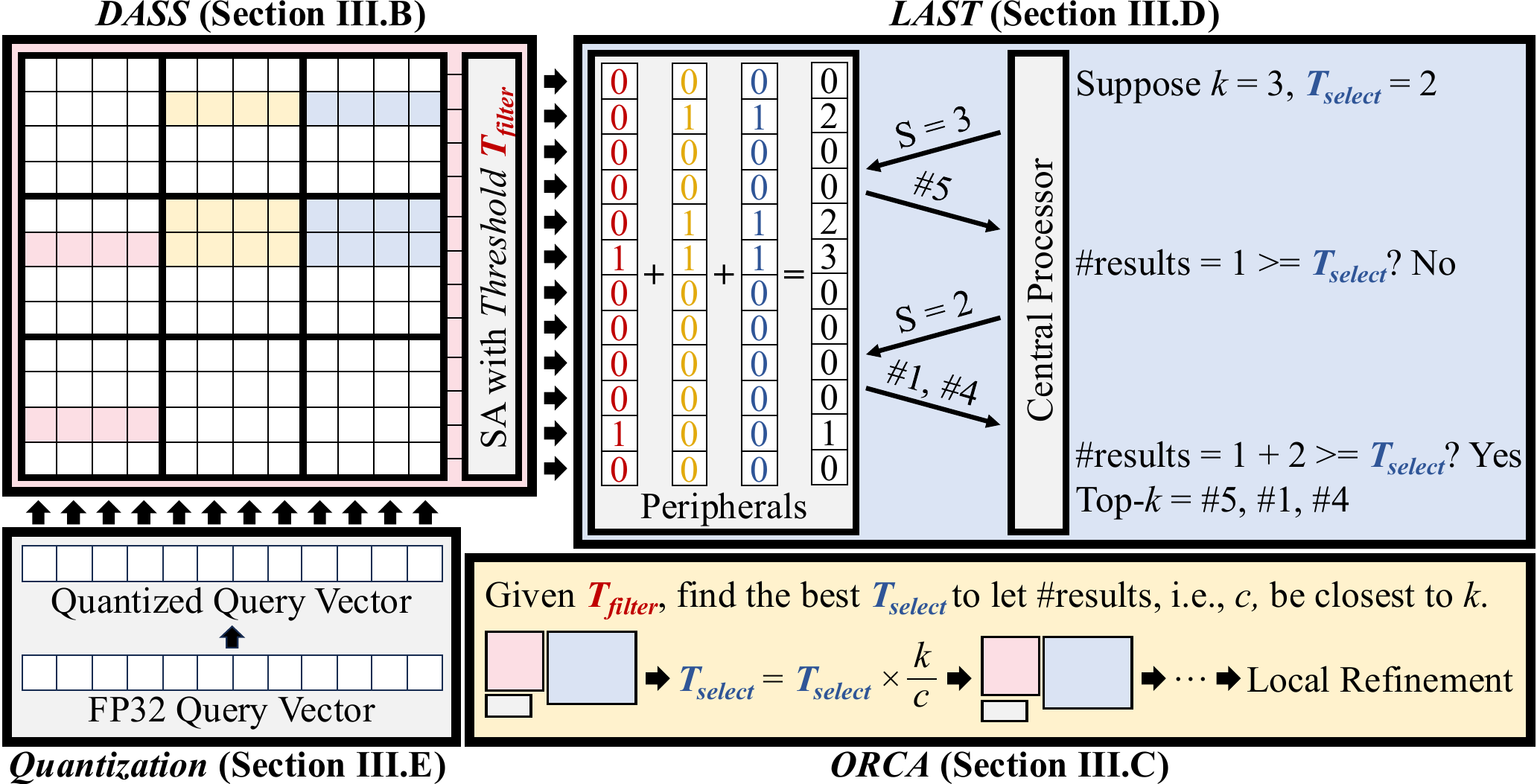}
    \caption{Overview of CAIMS.}
    \label{fig:Overview}
\end{figure}

To tackle these identified challenges, we design CAIMS around three key components (see Fig.~\ref{fig:Overview}): \textit{Dataset-Assisted Scalable Search (DASS)} and \textit{Online Result Count Adaptation (ORCA)} to achieve accurate search, and \textit{Local Aggregation and Selective Transfer (LAST)} to improve search efficiency. 

\textit{DASS} handles the accuracy challenge by unifying matching criteria across arrays, allowing local match results to be aggregated into globally comparable coarse-grained similarity scores. Since score ties can prevent selecting exactly \textit{k} results and the effectiveness of \textit{DASS} depends on the query distribution, we propose \textit{ORCA} to dynamically tune the search parameters, thereby adapting to query distribution shifts.

\textit{LAST} addresses the efficiency challenge by performing local aggregation and coordinated global selection using lightweight peripherals with negligible hardware overhead. By transferring only the global selection results and necessary control signals, \textit{LAST} reduces data movement to $O(k+\frac{\#dimensions}{array\ width})$, which is independent of the number of vectors in the dataset.

\subsection{Dataset-Assisted Scalable Search (DASS)}
As a key component of CAIMS, \textit{DASS} enables effective local-data filtering and global-result reduction that preserves search accuracy. By applying \textit{threshold match} with a unified threshold $T_{filter}$ in every CAM array, \textit{DASS} gives every local match result the same similarity interpretation: whether the similarity between a candidate subvector and the query subvector from the same dimension partition exceeds $T_{filter}$. Aggregating the local match results thus produces a globally comparable, coarse-grained similarity score, as shown in Fig.~\ref{fig:LocalInformation}.

One challenge with the above approach is that coarse-grained similarity scores generally may not identify exactly \textit{k} results because multiple candidates may tie at the top-\textit{k} cut-off. Since arbitrarily discarding tied candidates can degrade search accuracy, we introduce another threshold, ${T_{select}}$, such that \textit{DASS} returns every candidate whose score is at least that of the ${T_{select}}^{th}$ ranked candidate, aiming to keep the average result count, \textit{c}, as close to \textit{k} as possible. Selecting appropriate $T_{filter}$ and $T_{select}$ values is key to maintaining $c \approx k$ and achieving high  top-\textit{k} similarity search accuracy.

\textit{DASS} jointly determines $T_{filter}$ and $T_{select}$  through offline calibration, as outlined in Algorithm~\ref{alg:DASS}. Since actual query vectors are unavailable during system setup, \textit{DASS} randomly samples \(N_{smp}\) candidate vectors from the dataset as pseudo queries, using the candidate distribution as an initial approximation of the query distribution. The number of candidate $T_{filter}$ and the respective values are decided by users based on the implementation of \textit{threshold match}, e.g., adjusting sensing time or reference voltage \cite{EVACAM}, and the respective constraints. In this work we assume 8 evenly spaced candidate $T_{filter}$.

For each candidate $T_{filter}$ value, \textit{DASS} initializes $T_{select}$ to \textit{k}. Since \textit{c} is monotonically nondecreasing with $T_{select}$ for a fixed $T_{filter}$ value, \textit{DASS} applies the proportional update $T_{select} = T_{select} \times \frac{k}{c}$ until a previously evaluated  $T_{select}$ value is reached. Subsequently, \textit{DASS} performs \textit{NeighborRefine} to identify the locally optimal $T_{select}$, i.e., the setting whose \textit{c} is closest to \textit{k}. To reduce the risk of overfitting to the calibration samples, \textit{NeighborRefine} additionally performs a stability check: both neighboring settings, $T_{select} - 1$ and $T_{select} + 1$, must also select approximately \textit{k} candidates. If the locally optimal setting is unstable, \textit{NeighborRefine} (pseudocode omitted for simplicity) searches the neighboring direction for a nearby stable setting. A candidate $T_{filter}$ value is discarded if no stable setting is found. Among the remaining stable pairs, \textit{DASS} returns the pair $(T_{filter}, T_{select})$ whose \textit{c} is closest to \textit{k}. If no stable pair exists, \textit{DASS} instead returns the pair whose \textit{c} is closest to \textit{k}.

\begin{algorithm}[t]
\caption{\textit{DASS} Offline Parameter Selection}
\label{alg:DASS}
\begin{algorithmic}[1]
    \STATE $\mathit{qry\_vec} \gets
    \mathrm{RandomSample}(\mathit{cand\_vec}, N_{\mathrm{smp}})$
    \STATE $T_{\mathrm{filter}} \gets \mathrm{UserDefinedCandidateTfilters}()$
    \STATE $T_{\mathrm{select}} \gets [k,\ldots,k]$

    \FOR{$i = 0$ to $|T_{\mathrm{filter}}|-1$}
        \WHILE{\textbf{true}}
            \STATE $c \gets
            \mathrm{AvgResultCount}(
            \mathit{qry\_vec},
            T_{\mathrm{filter}}[i],
            T_{\mathrm{select}}[i])$

            \STATE $T'_{\mathrm{select}} \gets
            \mathrm{Round}\left(
            T_{\mathrm{select}}[i] \times \frac{k}{c}
            \right)$

            \IF{$T'_{\mathrm{select}}$ has been evaluated}
                \STATE \textbf{break}
            \ENDIF

            \STATE $T_{\mathrm{select}}[i]
            \gets T'_{\mathrm{select}}$
        \ENDWHILE

        \STATE $T_{\mathrm{select}}[i]
        \gets \mathrm{NeighborRefine}(
        T_{\mathrm{filter}}[i],
        T_{\mathrm{select}}[i])$
    \ENDFOR

    \STATE \textbf{return}
    $\mathrm{BestPair}(T_{\mathrm{filter}},T_{\mathrm{select}})$
\end{algorithmic}
\end{algorithm}

\subsection{Online Result Count Adaptation (ORCA)}
Although \textit{DASS} can select approximately \textit{k} candidates for pseudo queries sampled from the vector dataset, runtime queries may follow a different distribution. This distribution discrepancy can result in deviation of \textit{c}, with lower or higher $c$ values reducing accuracy or efficiency, respectively.

We design \textit{ORCA}, as outlined in Algorithm~\ref{alg:ORCA}, to adapt to query distribution shifts at runtime. Since altering $T_{filter}$ online may be challenging, \textit{ORCA} tunes only $T_{select}$. Specifically, \textit{ORCA} collects online queries and monitors their \textit{c}. When the deviation, i.e., $\frac{|c-k|}{k}$, exceeds a predefined bound \textit{delta}, \textit{ORCA} follows the $T_{select}$ process in \textit{DASS}: updating $T_{select}$ by a factor of $\frac{k}{c}$ for fast convergence and applying \textit{NeighborRefine} for local refinement. With \textit{ORCA}, CAIMS adapts to runtime query-distribution shifts, and \textit{DASS} can be reapplied after substantial dataset updates if the original $T_{filter}$ no longer controls \textit{c} well.

\begin{algorithm}[t]
\caption{\textit{ORCA Online Parameter Tuning}}
\label{alg:ORCA}
\begin{algorithmic}[1]

\STATE $\mathit{qry\_vec} \gets
\mathrm{CollectOnlineQueries}(N_{\mathrm{online}})$

\STATE $c \gets
\mathrm{AvgResultCount}(
\mathit{qry\_vec},
T_{\mathrm{filter}},
T_{\mathrm{select}})$

\STATE $\mathit{dev} \gets
\frac{|c-k|}{k}$

\IF{$\mathit{dev} > \mathit{delta}$}

    \WHILE{\textbf{true}}
        \STATE $T'_{\mathrm{select}} \gets
        \mathrm{Round}\left(
        T_{\mathrm{select}} \times \frac{k}{c}
        \right)$

        \IF{$T'_{\mathrm{select}}$ has been evaluated}
            \STATE \textbf{break}
        \ENDIF

        \STATE $T_{\mathrm{select}}
        \gets T'_{\mathrm{select}}$

        \STATE $c \gets
        \mathrm{AvgResultCount}(
        \mathit{qry\_vec},
        T_{\mathrm{filter}},
        T_{\mathrm{select}})$
    \ENDWHILE

    \STATE $T_{\mathrm{select}}
    \gets \mathrm{NeighborRefine}(
    T_{\mathrm{filter}},
    T_{\mathrm{select}})$

\ENDIF

\STATE \textbf{return} $T_{\mathrm{select}}$

\end{algorithmic}
\end{algorithm}

\subsection{Local Aggregation and Selective Transfer (LAST)}

To minimize data movement for result reduction, we design \textit{LAST} to perform local aggregation and coordinated global selection, thereby transferring only the global selection results and necessary control signals. Fig.~\ref{fig:Architecture}(a) illustrates the general CAIMS hardware architecture supporting \textit{LAST}. By mapping subvectors belonging to the same candidate to different arrays in the same processing element (PE), local binary match results can be aggregated to coarse-grained similarity scores at the PE layer. After obtaining the scores, \textit{LAST} performs coordinated global selection. The central processor first broadcasts the maximum possible score \textit{Max} = $\lceil \frac{\#dimensions}{array\ width} \rceil$ to all PEs, and each PE returns the candidates whose score equals \textit{Max}. The central processor cumulatively counts the returned results and terminates once at least $T_{select}$ results have been collected; otherwise, it broadcasts \textit{Max}-1 and repeats the process. By adopting \textit{LAST}, only the scores and the global selection results are transferred across the memory hierarchy, eliminating unnecessary data movement. Moreover, the distributed local processing design allows the reduction latency to remain constant and depends only on the value $k$ as the dataset size increases.

The Local Peripherals in Fig.~\ref{fig:Architecture}(a) support \textit{LAST} with low-bit-width accumulators for aggregation and equality comparators for selection. Since these operations are faster than CAM search, CAIMS provisions only enough peripheral resources to keep reduction off the critical path, minimizing hardware overhead without sacrificing throughput. The General Peripherals handle communication and address decoding and are standard IMS components rather than CAIMS-specific additions.

\subsection{Case Study: RAG Retrieval on 3D NAND Flash}
\begin{figure}[t]
    \centering
    \includegraphics[width=1.0\linewidth]{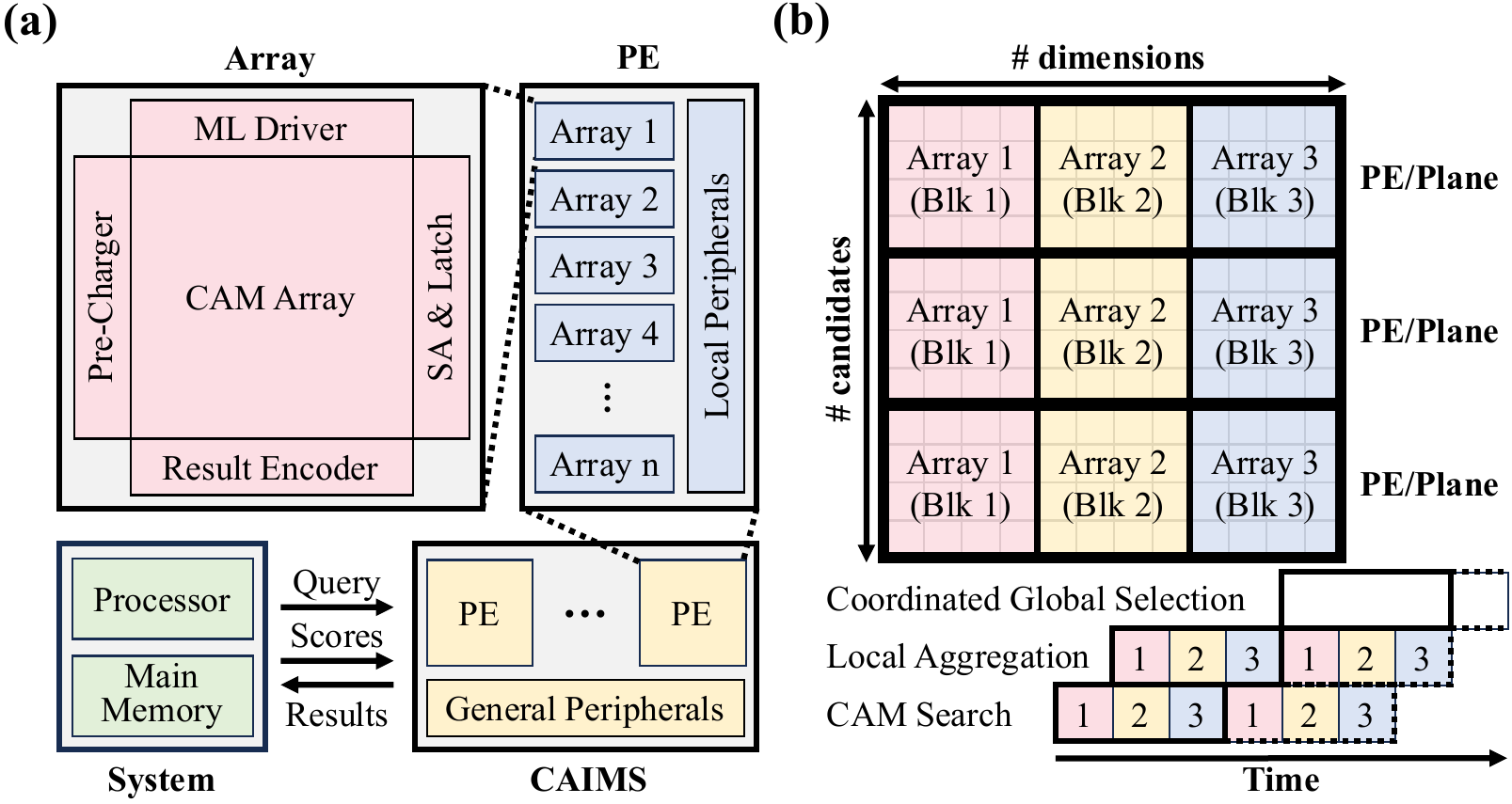}
    \caption{(a) CAIMS hardware architecture. (b) Pipeline for SSD implementation.}
    \label{fig:Architecture}
\end{figure}

We implement CAIMS with a 3D NAND Flash model and evaluate it using RAG retrieval. CAM arrays are mapped onto NAND blocks, PEs onto planes, and the central processor onto the SSD controller, with CAIMS spanning multiple chips. Fig.~\ref{fig:Architecture}(b) illustrates the execution pipeline. Since blocks within a plane share peripheral circuitry, CAIMS searches them sequentially while aggregating results from the previous search. Once coarse-grained similarity scores are available, the SSD controller initiates coordinated global selection while the blocks and peripherals begin processing the next query.

Quantization is required to map full-precision vectors onto limited-precision NAND cells. While quantization has advanced rapidly in recent years, most target model compression and therefore aim to minimize the reconstruction error \cite{GPTQ}. However, value reconstruction is unnecessary when a vector dataset is used solely for top-\textit{k} similarity search; instead, quantization should preserve the relative similarities among vectors. Conventional quantization schemes optimized for value reconstruction may hence be ill-suited for this purpose and can substantially degrade search accuracy. For instance, a recent work directly applying a scheme optimized for reconstruction reports 33\% search accuracy loss for 4-bit quantization \cite{4BitQuantization}.

We use \textit{Quantile Quantization} \cite{QuantileQuantization}, an information-theoretically optimal scheme \cite{QLoRA}, to quantize query and candidate vectors. Instead of minimizing reconstruction error, \textit{Quantile Quantization} determines quantization levels based on the quantile of the values in the input vector, ensuring approximate equal utilization of all quantization levels. By avoiding underutilization of quantization levels, similarity relationships between vectors become more distinguishable, which empirically makes top-\textit{k} selection significantly more accurate.

\section{evaluation}
Since CAIMS uses coarse-grained similarity scores for top-\textit{k} selection, trading accuracy for high search efficiency, we want to measure whether CAIMS improves throughput and energy efficiency while maintaining comparable accuracy. In addition, as CAIMS does not guarantee selecting exactly k candidates, whether \textit{DASS} and \textit{ORCA} can control \textit{c} is also important. In this section, we analyze these two aspects by implementing CAIMS with 3D NAND Flash model and evaluating on RAG retrieval. The experimental configurations are as follows:

\textbf{Baseline Implementations:} All datasets are encoded as 768 dimensional vectors with \textit{BAAI/bge-base-en-v1.5} \cite{BGE}. We implement CPU-based brute-force search for theoretical-optimal accuracy, and CPU-based IVF with FAISS \cite{FAISS} as a practical option. Both implementations set batch size to 16, following the throughput-optimized setting of \cite{IKS}. The IVF parameters are shown in Table~\ref{tab:Dataset}. For hardware acceleration, we compete against two representative 3D NAND Flash accelerators, TVLSI2026 \cite{TVLSI26} and D-NOVA \cite{D-NOVA}, which report superior performance over other baselines such as REIS \cite{REIS, Macronix1, Macronix3, Macronix4}, with statistics provided in the source papers.

\textbf{Datasets:} Since CAIMS is distribution-dependent, we evaluate CAIMS on six datasets of varying sizes from BEIR \cite{BEIR} and DPR \cite{DPR}, with detailed information summarized in Table~\ref{tab:Dataset}.

\begin{table}[t]
\centering
\caption{detailed dataset and ivf information}
\label{tab:Dataset}
\renewcommand{\arraystretch}{1.1}
\begin{tabular}{lrrr}
\hline
\textbf{Dataset} & \textbf{\#queries} & \textbf{\#candidates} & \textbf{IVF(\textit{n\_list} / \textit{n\_probe})} \\
\hline
BEIR-SciFact & 300    & 5,183 & 64 / 16 \\
BEIR-SciDocs & 1,000  & 25,657 & 128 / 16 \\
BEIR-Quora   & 10,000 & 522,931 & 512 / 32 \\
BEIR-NQ      & 3,452  & 2,681,468 & 1024 / 64 \\
DPR-NQ       & 3,610  & 21,015,324 & 4096 / 256 \\
DPR-TQA      & 11,313 & 21,015,324 & 4096 / 256 \\
\hline
\end{tabular}
\end{table}

\begin{table}[t]
\centering
\caption{system configuration}
\label{tab:Configuration}
\renewcommand{\arraystretch}{1.1}
\begin{tabular}{ll}
\hline
\textbf{Device} & \textbf{Description} \\
\hline
CPU  & Intel Xeon Gold 6530, SMT on \\
DRAM & 8 $\times$ 64GB DDR5-4800 \\
SSD &
\begin{tabular}[c]{@{}l@{}}
8 channels, 16 chips/channel, 2 planes/chip, 128 Kb/page, \\
5.8 pJ/bit transfer energy \cite{TVLSI26}, 25W power budget
\end{tabular} \\
\hline
\end{tabular}
\end{table}

\begin{table*}[t]
\centering
\caption{overall performance (hitrate@100/query per second/energy per query)}
\label{tab:Performance}
\renewcommand{\arraystretch}{1.1}

\begin{tabular}{
l
r@{ / }r@{ / }r
r@{ / }r@{ / }r
r@{ / }r@{ / }r
r@{ / }r@{ / }r
}
\hline
\textbf{Dataset}
& \multicolumn{3}{c}{\textbf{CPU-Based Brute-Force Search}}
& \multicolumn{3}{c}{\textbf{CPU-Based IVF}}
& \multicolumn{3}{c}{\textbf{TVLSI2026 \cite{TVLSI26}}}
& \multicolumn{3}{c}{\textbf{CAIMS}} \\
\hline

BEIR-SciFact
& \textbf{97.0\%} & 11063.8 & 21.2\,mJ
& 94.7\% & \textbf{64840.8} & 3.9\,mJ
& \multicolumn{3}{c}{Not Provided}
& 95.0\% & 625.0 & \textbf{0.2\,mJ} \\

BEIR-SciDocs
& \textbf{91.5\%} & 4285.1 & 69.6\,mJ
& 91.1\% & \textbf{27656.3} & 9.5\,mJ
& \multicolumn{3}{c}{Not Provided}
& 88.8\% & 625.0 & \textbf{0.2\,mJ} \\

BEIR-Quora
& \textbf{99.9\%} & 158.8 & 1884.1\,mJ
& 99.6\% & \textbf{1282.3} & 247.6\,mJ
& \multicolumn{3}{c}{Not Provided}
& \textbf{99.9\%} & 625.0 & \textbf{0.7\,mJ} \\

BEIR-NQ
& \textbf{94.0\%} & 31.0 & 9657.5\,mJ
& 91.7\% & 238.8 & 1343.1\,mJ
& \multicolumn{3}{c}{Not Provided}
& 92.0\% & \textbf{625.0} & \textbf{3.5\,mJ} \\

DPR-NQ
& \textbf{89.1\%} & 3.6 & 83895.7\,mJ
& 88.7\% & 28.0 & 11602.8\,mJ
& 85.8\% & 169.2 & 609.8\,mJ
& 88.2\% & \textbf{625.0} & \textbf{27.2\,mJ} \\

DPR-TQA
& \textbf{86.2\%} & 3.9 & 77785.0\,mJ
& 85.9\% & 30.9 & 10545.3\,mJ
& 84.5\% & 62.1 & 643.7\,mJ
& 84.0\% & \textbf{625.0} & \textbf{27.2\,mJ} \\

\hline
\end{tabular}
\end{table*}

\textbf{Evaluation Metrics:}
Following TVLSI2026, we evaluate accuracy with HitRate@100 and efficiency with throughput and energy per query. CAIMS and software baselines respectively use squared Euclidean distance and cosine similarity, equivalent for ranking $L_2$-normalized full-precision vectors.

\textbf{System Configuration:}
Table~\ref{tab:Configuration} lists the detailed compute and memory configurations. We execute the algorithmic baselines on the listed server-class CPU and DRAM, while modeling CAIMS with the listed TLC SSD \cite{PM9A3}, which REIS \cite{REIS} characterizes as a ``low-cost commercial SSD.''

For CAIMS, we configure 48 wordline (WL) layers with complementary encoding using serially coupled cells, corresponding to a CAM array width of 24 and 32 partitions for each 768-dimensional vector. Candidate subvectors are mapped vertically along NAND strings, with each query dimension applied to its corresponding WL. These configurations follow a fabricated design that demonstrates robust search accuracy \cite{48Layers}. Vectors are quantized to 3-bit precision and stored in TLC cells, whose feasibility for 3D NAND Flash-based similarity search has been demonstrated in prior work \cite{TLC}.

\textbf{Latency and Energy Modeling:}
We model search latency as 50 \ensuremath{\mu}\text{s} based on the reported 30--50 \ensuremath{\mu}\text{s} range for 3D NAND \textit{threshold match} \cite{48Layers}. We conservatively estimate search energy as 4.79~\ensuremath{\mu}\text{J} per block, twice the reported 128-Kb page read energy \cite{AiF}, since reads apply $V_{\mathrm{Read}}$ and $V_{\mathrm{Pass}}$ to the selected and remaining WLs, respectively, while searches apply query-dependent voltages across WLs. Total energy is modeled as
\[
E_{\mathrm{total}}
=
E_{\mathrm{search}}
+
E_{\mathrm{reduction}}
+
E_{\mathrm{transfer}}
+
E_{\mathrm{Controller}}.
\]
Since (1) reduction energy is negligible compared with search WL activation \cite{D-NOVA}, (2) transfer energy is orders of magnitude smaller than search energy, and (3) SSD-controller operations consist primarily of lightweight counting and control, we conservatively approximate the total energy as $E_{\mathrm{total}} = 1.1E_{\mathrm{search}}$.

\subsection{Overall Performance}
Table~\ref{tab:Performance} compares the overall performance of CAIMS with different baselines. CPU-based IVF trades a slight accuracy loss for significantly higher throughput and energy efficiency than CPU-based brute-force search, making it a practical software baseline. However, its throughput decreases rapidly with dataset size, from 64.8k QPS on BEIR-SciFact to 28.0 QPS on DPR-NQ. In contrast, CAIMS throughput is determined primarily by the fixed 3D NAND search latency within its parallel search capacity. With the listed SSD, CAIMS searches up to 33.6M ($8\times16\times2\times128$k) candidate vectors in parallel and sustains 625.0 QPS through 32 sequential subvector searches per 768-dimensional vector. Although CPU-based IVF is preferable for small datasets, CAIMS achieves up to $22.3\times(426.6\times)/10.1\times(23.7\times)$ improvements in throughput (energy) with comparable accuracy over CPU-based IVF (on average -0.6\%)/TVLSI2026 (on average +0.9\%). Beyond 33.6M candidates, multiple search waves reduce throughput to $625.0/\left\lceil\#\mathrm{candidates}/33.6\mathrm{M}\right\rceil$ QPS. Note that the effective CAM array width can be adjusted by selecting which WLs participate in each comparison. A narrower width provides finer-grained similarity scores, improving accuracy, whereas a wider width reduces search cycles, enhancing efficiency.

TVLSI2026 also exploits in-storage parallelism but is query-sensitive, requires substantial in-block modification, and assumes an optimistic array width of 768. CAIMS uses only low-bit-width accumulators and equality comparators per plane, yet achieves higher throughput, energy efficiency, and average accuracy. Although D-NOVA reports only normalized results, the efficient DTS242 configuration requires 60 search cycles and critical-path sorting for 768-dimensional vectors on a 48-layer SSD. CAIMS needs only 32 cycles and pipelines reduction with search, yielding at least $1.88\times$ higher throughput. As search dominates energy consumption, CAIMS is also expected to be more energy-efficient than D-NOVA.

\subsection{Control of \textit{c}}
\begin{figure}[t]
    \centering
    \includegraphics[width=1.0\linewidth]{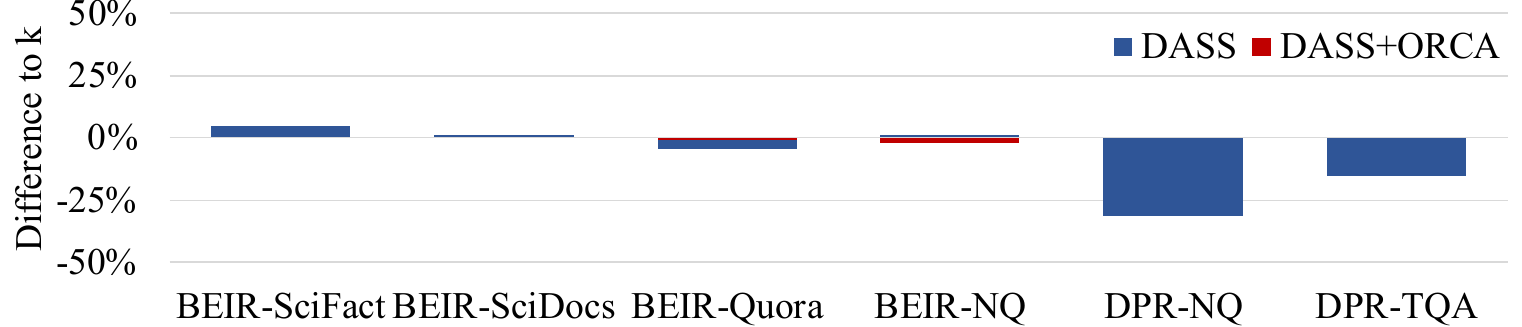}
    \caption{Effectiveness of \textit{DASS} and \textit{ORCA} in controlling \textit{c} with \textit{k}=100.}
    \label{fig:Control}
\end{figure}

As query distribution is stationary within each dataset, we directly apply \textit{ORCA} once for each dataset to validate the adaptation without setting \textit{delta}. Figure~\ref{fig:Control} presents the relationship between \textit{c} and \textit{k} in six datasets. Experimental results show that the offline parameter selection of \textit{DASS} can already control \textit{c} well for small- and medium-sized datasets, whereas \textit{ORCA} complements it on larger datasets. CAIMS controls the average difference to \textit{k} to be 0.7 with merely 300 online queries, indicating that CAIMS selects on average \textit{k} candidates for each query. Smaller \textit{k} also follows this trend, though we omit the statistics here due to page limitation.

\subsection{Design Feasibility}
We explicitly consider (1) peripheral area overhead and (2) the SSD power constraint when evaluating design feasibility. LAST requires only lightweight accumulation and equality-comparison logic at each plane, similar to the peripherals adopted by D-NOVA, which incur less than $1.5\%$ plane-area overhead, suggesting that the additional hardware required by CAIMS is modest. For power, the modeled SSD has a $25\,\mathrm{W}$ form-factor power limit \cite{PM9A3}. Under the evaluated DPR workloads, CAIMS consumes at most $17.0\,\mathrm{W}$ with conservative estimation, whereas TVLSI2026 would require approximately $100\,\mathrm{W}$ and $40\,\mathrm{W}$ for DPR-NQ and DPR-TQA, respectively. We note that simultaneously activating all planes in the SSD may exceed the power envelope. In such cases, power-aware scheduling would be required, which we omit in this work. Overall, these results indicate that CAIMS can potentially be implemented with commercial SSDs.

\section{conclusion}
IMS offers great potential for accelerating search-based data-intensive applications. However, existing cross-array designs remain constrained along the entry or dimension axis, failing to scale dataset size and vector dimensionality simultaneously. This work presents CAIMS, a cross-array IMS framework that enables fully scalable and efficient top-\textit{k} similarity search while preserving accuracy. Evaluated using RAG retrieval on a 3D NAND Flash model, CAIMS improves energy efficiency across all evaluated datasets and throughput on large datasets. These results establish CAIMS as a promising framework for large-scale IMS. We envision CAIMS broadening the applicability of CAM-based search to various workloads, as dataset scale is no longer bounded by array capacity.

\section{content generated by ai}
Claude and ChatGPT assisted with code development, with all generated content reviewed and verified by the authors.

\newpage

\bibliographystyle{IEEEtran}
\bibliography{references}

\end{document}